\documentclass[iop]{emulateapj}

\usepackage{natbib}

\newcommand{\B}{\rule[-1.2ex]{0pt}{0pt}}

\begin{document}


\title{C{\small IV} and C{\small III]} reverberation mapping of the luminous quasar PG~1247+267  \\}


\author{D. Trevese$^1$, M. Perna $^2$, F. Vagnetti$^3$, F.G. Saturni$^{1,4}$, and M. Dadina$^5$}
         
\affil{$^1$Dipartimento di Fisica, Universit\`a di Roma La Sapienza, Piazzale Aldo Moro, 5, 00185 Roma, Italy\\
$^2$ Dipartimento di Fisica e Astronomia, Universit\`a di Bologna, Viale Berti Pichat 6/2, 40127 Bologna, Italy\\
$^3$ Dipartimento di Fisica, Universit\`a di Roma Tor Vergata, Via della Ricerca Scientifica 1, 00133 Roma, Italy\\
$^4$ European Southern Observatory, Karl-Schwarzschild-Strasse 2, 85748 Garching bei M\"unchen, Germany\\
$^5$ INAF-IASF Bologna, Via Gobetti 101, 40129 Bologna, Italy}



\begin{abstract}
So far the masses of about 50 active galactic nuclei have been measured through the reverberation mapping technique (RM). Most measurements have been performed for objects of moderate luminosity and redshift, based on H$\beta$, which is also used to calibrate the scaling relation which allows single-epoch (SE) mass determination based on AGN luminosity and the width of different emission lines. The SE mass obtained from C{\scriptsize IV}$(1549 {\rm\AA})$ line shows a large spread around mean values,  due to complex structure and gas dynamics of the relevant emission region. Direct RM measures of C{\scriptsize IV} exist for only 6 AGNs of low luminosity and redshift, and only one luminous quasar \citep{Kasp07}. We have collected since 2003 photometric and spectroscopic observations of PG1247+267, the most luminous quasar ever analyzed for RM. We provide light curves for the continuum and for C{\scriptsize IV}$(1549 {\rm\AA})$ and C{\scriptsize III]}$(1909{\rm\AA})$, and measures of the reverberation time lags based on the SPEAR method \citep{Zu11}. The sizes of the line emission regions are in a ratio $R_{CIII]}/R_{CIV}\sim 2$, similar to the case of Seyfert galaxies, indicating for the first time a similar ionization stratification in a luminous quasar and low luminosity nuclei. Due to relatively small broad line region size and relatively narrow line widths, we estimate a small mass and an anomalously high Eddington ratio. We discuss the possibility that either the shape of the emission region or an amplification of the luminosity caused by gravitational lensing may be in part responsible of the result.  
\end{abstract}



\section{Introduction}
Reverberation mapping (RM) has played a crucial role in the study of the structure of active galactic nuclei (AGN). Spectroscopic monitoring  in the UV/optical band permits to measure emission line flux changes,  representing  the "echo" of the far UV ionising  continuum variations,  which, in turn are closely related with the observed near UV continuum variations. The delay $\tau_l$ of the echo, i.e. of line variation with respect to continuum changes, is measured through the continuum-line cross-correlation and provides the luminosity-weighted average distance, $R=c \cdot \tau_l$  of the line-emitting region from the (point-like) continuum source, where $c$ is the speed of light \citep{Blan82,Pete93}. Until 1999,  17 AGNs with $\lambda L_{\lambda}(5100 {\rm\AA}) \lesssim 1.5 \times10^{44}$ erg  s$^{-1}$ were studied \citep[see][and refs therein]{Wand99} with the result of measuring the size of their  broad line regions (BLR) and demonstrating a  stratification of ionisation, with the higher ionisation lines responding more rapidly to continuum changes. Combining the size $R$, with a measure of the typical velocity $\Delta V$ of the emitting BLR clouds, assumed in Keplerian orbits, it is possible to derive a  virial estimate of the black hole mass $M_{BH}=f c \tau_l \Delta V^2/G$ of the central black hole, where  $G$ is the gravitational constant and $f$ is a scaling factor depending on the geometry of the BLR and the specific definition adopted for $\Delta V$ (see section 4.). The extension of these results by the addition of a sample of 17  quasars (QSO)  with luminosities $\lambda L_{\lambda}(5100 {\rm\AA})$ up to $\sim 6.5 \times10^{45}$ erg  s$^{-1}$, allowed \citet{Kasp00} to establish a size-luminosity  scaling relation of the type $R\propto L^{\gamma}$, in a luminosity range covering more than four decades  \citep{Kasp05,Bent06,Bent09}.
 This relation can be used to estimate the BH mass on the basis of $\Delta V$ and $L$ measured from single epoch (SE) spectra \citep{Vest02,McLu02}, opening the possibility of estimating the BH mass of thousands QSOs/AGNs, analysing their luminosity function at different redshifts,  and following the BH-galaxy co-evolution in cosmic time \citep{Shen12}. 
The widths of different lines, H$\beta$, C{\scriptsize IV}, Mg{\scriptsize II}, are used depending on redshift and  wavelength range of
optical/IR ground based observations. However,  the scaling relations for C {\scriptsize IV} and Mg{\scriptsize II}   \citep{McLu02,Vest06,McGi08} are not obtained  from the few direct RM measures of these lines, but are calibrated on the mass scale based on H$\beta$ time lags, which represent the majority of RM measures to date.
The latter are presently limited to objects with $\lambda L_{\lambda}(5100 {\rm\AA}) \lesssim$ 10$^{46}$ erg s$^{-1}$ and $z<0.3$ \citep[][and refs therein]{Bent13}.
According to \citet{Netz03},  the largest black hole masses deduced from these extrapolations, occurring in objects with the highest luminosities,  would exceed 10$^{10} M_{\odot}$, and,  if converted to host galaxy mass and luminosity through the statistical relation among the black hole mass, galaxy bulge mass  and stellar velocity dispersion,  would imply galactic bulge masses  $M_{bulge}\gtrsim10^{13} M_{\odot}$ and stellar velocity dispersions exceeding 700 km/s which have never been observed, suggesting that either   the $M_{BH}$-$M_{bulge}$ correlations observed in the local universe are different at higher redshift, or  the observed size-luminosity relationship in low-luminosity AGNs does not extend to very high luminosity.
 \citet{Vest04}  pointed out that the space density of such luminous quasars is so low that  their local absence does not mean they don't exist. In any case, exploring the validity or failure of the size-luminosity scaling relation is of crucial importance not only to understand the physical conditions in the most luminous QSOs, but also because 
 most of the AGN mass estimates are based this unconfirmed extrapolation,  which could lead to uncertain or biased conclusions on the evolution of the AGN mass function in cosmic time.

To measure the  BRL size and BH mass of luminous QSOs, in 2003 we started a monitoring campaign of four high luminosity ($L\gtrsim 5 \times 10^{46}$erg s$^{-1}$) and intermediate redshift ($2<z<4$) objects with the Copernico  1.82 m telescope in Asiago (Italy) and the Cassini 1.52 m telescope in Loiano (Italy). Some results  on the QSOs PG~1634+706, with $z=1.337$ and PG~1247+267, with $z=2.048$, showing the detectability of the emission line variations, were published in \citet{Trev07}. A study of broad absorption line variability of  the luminous quasar APM~08279+5255 \citep{Trev13,Satu14} and preliminary results on RM for PG~1247+267 \citep{Pern14} were also presented.

At $z\gtrsim 2$ H$\beta$, is no longer observable in the optical band and reverberation can be observed for C {\scriptsize III]}  
($\lambda 1909$ {\rm\AA}) and  C {\scriptsize IV}($\lambda 1549$ {\rm\AA}) lines.   Reverberation measurements of the   C {\scriptsize IV} line are available only for a handful of low luminosity ($\lambda L_{\lambda}(1350 {\rm\AA}) \lesssim 10^{44}$ erg s$^{-1}$) and low redshift ($z<0.06$) AGN observed in the ultraviolet  from space.  In addition, \citet{Kasp07}  presented the first results of a RM campaign started in 1999 with HET  11 m telescope \citep{Rams98} providing a first tentative mass estimate for S5 0836+071, a luminous ($\lambda L_{\lambda}(1350 {\rm\AA})= 1.12\pm 0.16 \times 10^{47}$ erg s$^{-1}$) QSO at $z=2.172$, based on C{\scriptsize IV} RM.
More recently, several studies discussed the unreliability of C {\scriptsize IV} based mass estimates, due to gas outflows strongly affecting the profile of this line \citep{Netz07,Sule07,Marz12,Denn12}.  The fact that there is no consensus about the scatter and possible biases  between C{\scriptsize IV}-based and H$\beta$-based  BH masses \citep{Gree10,Asse11,Runn13} further increases the importance of RM measure of the size of the emitting region to constrain wind models and eventually lead to a consistent picture which includes BH mass, gas outflow and possibly its feedback on the host galaxy. 

In this work we present the  C{\scriptsize IV}, C{\scriptsize III]} and continuum light curves obtained for PG~1247+267, and we estimate  the relevant time lags based on a  method proposed by \citet{Zu11}. We also analyze the shape of C{\scriptsize IV} and C{\scriptsize III]} lines,  and discuss the determination of the virial mass of the central BH, the corresponding value of the Eddington ratio an possible explanations of the anomalously high value found. The paper is organized as follows: in section 2 we describe observation and data reduction; in section 3 we discuss  the  estimate of time lags; in section 4 we discuss the mass estimates based on C{\scriptsize IV}, C{\scriptsize III]}  RM; in section 5 we draw the conclusions.

Throughout the paper we adopt the cosmology H$_o$=70 km s$^{-1}$, Mpc$^{-1}$,  $\Omega_m=0.3$, and $\Omega_{\Lambda}$=0.7.

\section{Observations and data reduction}
The majority of observations were carried out with the Faint Object Spectrograph \& Camera AFOSC at the Copernico 1.82 m telescope in Asiago (Italy).  We measured relative spectrophotometric variations by including a reference star in  a wide ($8''.44$) slit to avoid differential flux losses caused by atmospheric refraction. The reference star is the object at $\alpha=12 50 11.44$, $\delta=+26 33 32.1$ (J2000), with V=13.824 \citep{Pick10}. At each epoch, the observations consist of two consecutive exposures of $\sim1800$ s. Typical resolution is $\sim$15 {\rm\AA}~ in the spectral range 3500-8500 {\rm\AA}.  Details are described in \citet{Trev07}. The QSO and reference star uncalibrated spectra, $Q(\lambda)$ and $S(\lambda)$ respectively, are extracted by the standard IRAF\footnote{IRAF is distributed by the National Optical Astronomy Observatories,
which are operated by the Association of Universities for Research
in Astronomy, Inc., under cooperative agreement with the National
Science Foundation.} procedures, and the ratio $\mu^{(k)}=Q^{(k)}(\lambda)/S^{(k)}(\lambda)$ is computed for
 each exposure $k=1,2$. This quantity is independent of  extinction variations  and allows to reject inconsistent exposure pairs whenever $|\mu^{(2)}/\mu^{(1)}-1|$ averaged  over 500 {\rm\AA}~  exceeds 0.04.  This procedure also allows us to  compute the relative flux differences,  between the two exposures,  which are used to estimate the statistical errors on  continuum and emission line fluxes. At the $i$-th epoch $t_i$,  pairs of spectra of both the QSO and the reference star were co-added to compute the ratio $\mu_i(\lambda)=Q_i(\lambda)/S_i(\lambda)=(Q^{(1)}+Q^{(2)})/(S^{(1)}+S^{(2)})$. Data separated by less than one day are combined into a  single epoch data point.
\begin{figure}
\includegraphics[scale=0.44]{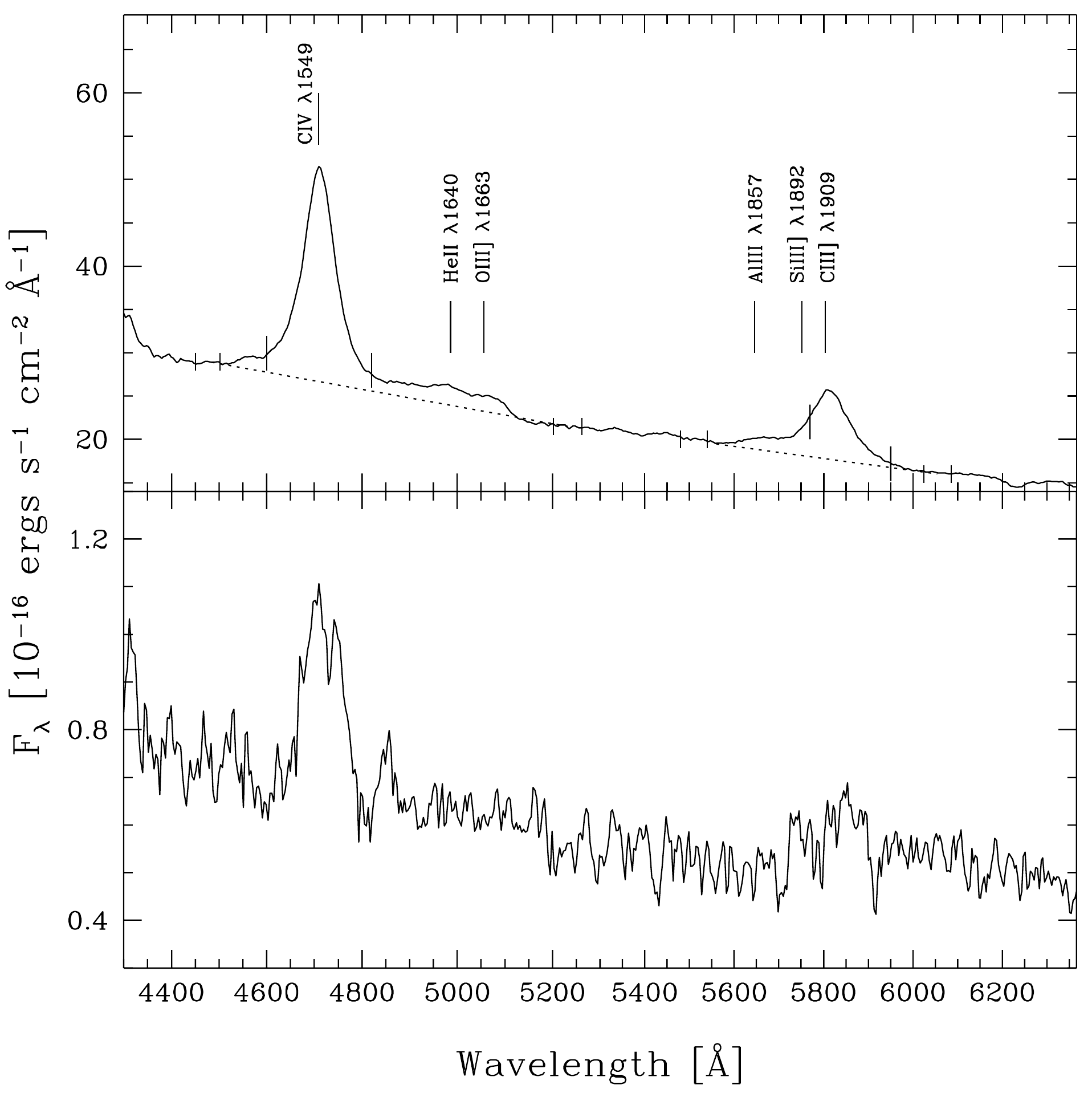}
\caption{ {\it Upper panel: } average spectrum of PG~1247+267 from our observations. Spectral ranges selected for the determination of the local continua (short ticks) and for the line flux (long ticks) are marked on the spectrum. Dotted lines represent the interpolated local continuum. {\it Lower panel:}  r.m.s. spectrum as defined in  \citet{Pete98}.
}
\label{fig:spec}
\end{figure}

The  flux-calibrated spectrum of the star $F^S(\lambda)$ was obtained at a single reference epoch, and the calibrated quasar spectra were obtained for each epoch as $F^Q_i(\lambda)=\mu_i(\lambda)F^S(\lambda)$.  We stress that the spectra are independent of extinction changes and detector response, thus spectral variations are also independent of telescope, detector and calibration. This allows us to include 4 spectra taken at the 1.5 m Cassini telescope of the Loiano Observatory, with the BFOSC camera. At each epoch two exposures of  2700 s  were taken,  with  about the same resolution of AFOSC spectra.

Although the following reverberation analysis is independent of the absolute calibration, we have now revised the calibration of the spectra for possible future uses. This has been done by multiplying the calibrated spectrum of the reference star by a constant factor which makes its V magnidute, as computed from the spectrum adopting the \citet{Bess90} filter profile, equal to  V=13.824 as given by  \citet{Pick10}.

Figure \ref{fig:spec} shows the flux-calibrated average spectrum and the r.m.s. spectrum of PG~1247+267. A spectral decomposition of the Al{\scriptsize III} +Si {\scriptsize III]} + C{\scriptsize III]} blend indicates a Si{\scriptsize III]}/C{\scriptsize III]}  flux ratio $\lesssim$ 0.3. This  is consistent with the corresponding ratio reported in table 2 of  \citet{Bach04}, since PG~1247+267 belongs to class B of \citet{Sule02} on the basis of the FWHM of the broad component of its H$\beta$ line  ($7460\pm 220$ km s$^{-1}$), as measured by \citet{Mcin99}. The integration limits adopted further reduces by a factor $\sim$2  the contribution to the computed C{\scriptsize III]} flux of  Si{\scriptsize III]}, which  therefore has been neglected. Similarly, the adopted continuum and integration limits should avoid the contamination of C{\scriptsize IV} flux from He{\scriptsize II}(1640{\rm\AA}) and O{\scriptsize III]}(1663{\rm\AA}) emission.
The lower panel of Figure \ref{fig:spec} shows the r.m.s. spectrum. As expected, it appears more noisy than the average spectrum \citep[cf.][]{Denn12}. Nonetheless, in the case of the stronger line C{\scriptsize IV} the average and r.m.s. profiles appear similar (see sect. 4).

Line fluxes are computed as $f_l = \int_{\lambda_1}^{\lambda_2}[F^{(Q)}(\lambda) - c^{int}(\lambda)] d \lambda$, where $c^{int}(\lambda)$ is the linear interpolation through the continua at shorter and longer wavelengths of each line, $\lambda_{short}$ and $\lambda_{long}$,    indicated in Figure \ref{fig:spec},  and the  extremes of integration $\lambda_1$ and ${\lambda_2}$, chosen to optimise the $f_l$ signal to noise ratio, not necessarily coincide with $\lambda_{short}$ and $\lambda_{long}$ (see Trevese et al. 2007).


\begin{table*}[t]
\caption{Variability measurements for PG 1247+267. \label{tab-journal} }     
\begin{center}                          
\scalebox{1.0}{
\begin{tabular}{l l cccrccc}        
\tableline               
\hline                        
Date& MJD& Telescope$^a$ &$\Delta V$&$\Delta R$&$\delta R~~~~~~~$&$F^{cont}_{\lambda}(5300 {\rm\AA}) $&$f_{CIII]}$ & $f_{CIV}$\\
		&		&		&\multicolumn{3}{c} \hrulefill&           &\multicolumn{2}{c} \hrulefill \\
		&		&        &\multicolumn{2}{c} {~~~~~~~~~~~~~~~~~~~magnitudes}&                &[$10^{-15}$ erg cm$^{-2}$ s$^{-1}$ {\rm\AA}$^{-1}$] &\multicolumn{2}{c}{[$10^{-14}$ erg cm$^{-2}$s$^{-1}$ ]}\\
\hline

03-01-25  & 52665.3 & L & $1.660\pm0.003$ & -             &$ -0.122    \pm 0.017$  & -                & -                 &  -             \\
03-02-23  & 52694.5 & L & $1.665\pm0.003$ & -             &$ -0.117    \pm 0.017$  & -                & -                 &  -             \\
03-04-01  & 52733.4 & L & $1.696\pm0.003$ & $1.80\pm0.01$ &$ -0.090    \pm 0.023$  & -                & -                 &  -             \\
03-05-09  & 52769.3 & A &         -       & $1.83\pm0.01$ &$ -0.034    \pm 0.023$  & $2.34\pm0.02$    &  $8.2\pm0.2$      & $25.6\pm0.4$   \\
04-01-13  & 53017.6 & A &         -       & $1.85\pm0.01$ &$ -0.014    \pm 0.023$  & $2.09\pm0.02$    &  $7.0\pm0.1$      & $23.0\pm0.3$   \\
04-01-14  & 53018.6 & A &         -       & $1.82\pm0.01$ &$ -0.048    \pm 0.023$  &   -              &   -               &  -             \\ 
04-02-12  & 53047.5 & A &         -       & $1.88\pm0.01$ &$  0.013    \pm 0.023$  & $2.15\pm0.02$    &  $7.3\pm0.2$      & $23.6\pm0.4$   \\
04-03-18  & 53083.0 & L & $1.758\pm0.003$ & $1.83\pm0.01$ &$ -0.035    \pm 0.023$  & -                & -                 & -              \\
04-05-07  & 53133.9 & L & $1.780\pm0.003$ & $1.87\pm0.01$ &$  0.004    \pm 0.023$  & -                & -                 & -              \\
05-01-17  & 53388.6 & A &         -       & $1.86\pm0.01$ &$ -0.006    \pm 0.023$  & $2.11\pm0.02$    &  $7.7\pm0.2$      & $22.8\pm0.3$   \\
05-03-10  & 53439.6 & A &         -       & $1.87\pm0.01$ &$  0.004    \pm 0.023$  & $2.09\pm0.02$    &  $7.6\pm0.2$      & $23.0\pm0.3$   \\
05-04-26  & 53487.4 & L & $1.762\pm0.003$ & $1.84\pm0.01$ &$ -0.024    \pm 0.023$  & -                & -                 & -              \\
05-05-02  & 53493.5 & L & $1.775\pm0.003$ & $1.87\pm0.01$ &$  0.002    \pm 0.023$  & -                & -                 & -              \\
05-05-13  & 53503.5 & A &         -       &        -      &		- ~~~~~~~        & $2.15\pm0.02$    &  $7.7\pm0.2$      & $22.9\pm0.4$   \\
06-03-31  & 53826.4 & L & $1.771\pm0.003$ & $1.86\pm0.01$ &$ -0.010    \pm 0.017$  & -                & -                 & -              \\
06-04-26  & 53852.5 & A &         -       & $1.86\pm0.01$ &$ -0.009    \pm 0.023$  & $2.15\pm0.02$    &  $7.5\pm0.2$      & $21.4\pm0.3$   \\
06-05-30  & 53886.5 & A &         -       & $1.89\pm0.01$ &$  0.022    \pm 0.023$  & $2.10\pm0.02$    &  $6.8\pm0.1$      & $22.0\pm0.3$   \\
08-04-02  & 54559.4 & A &         -       & $1.89\pm0.01$ &$  0.025    \pm 0.023$  & $2.04\pm0.02$    &  $7.6\pm0.2$      & $22.5\pm0.3$   \\
08-12-23  & 54823.7 & A &         -       & $1.88\pm0.01$ &$  0.012    \pm 0.023$  & $2.18\pm0.02$    &  $6.7\pm0.1$      & $23.2\pm0.3$   \\
09-03-23  & 54915.7 & A &         -       & $1.87\pm0.01$ &$  0.005    \pm 0.023$  & $2.17\pm0.02$    &  $7.0\pm0.1$      & $22.7\pm0.3$   \\
09-05-27  & 54979.5 & A &         -       & $1.87\pm0.01$ &$  0.005    \pm 0.023$  & $2.16\pm0.02$    &  $6.9\pm0.1$      & $23.0\pm0.3$   \\
11-04-01  & 55652.5 & A &         -       &         -     &		- ~~~~~~~         & $2.00\pm0.02$    &  $7.6\pm0.2$      & $25.3\pm0.4$   \\
12-02-27  & 55985.5 & A &         -       & $1.96\pm0.01$ &$  0.093    \pm 0.023$  & $2.13\pm0.02$    &  $6.8\pm0.1$      & $22.5\pm0.3$   \\
12-12-11  & 56272.7 & L &         -       & $1.78\pm0.01$ &$ -0.092    \pm 0.023$  & $2.21\pm0.02$    &  $7.2\pm0.2$      & $24.4\pm0.4$   \\
13-04-13  & 56396.4 & L &         -       & $1.74\pm0.01$ &$ -0.125    \pm 0.023$  & $2.37\pm0.02$    &  $6.9\pm0.1$      & $23.4\pm0.4$   \\
24-03-14  & 56741.6 & L &         -       & $1.80\pm0.01$ &$ -0.073    \pm 0.023$  & $2.33\pm0.02$    &  $6.4\pm0.1$      & $24.0\pm0.4$   \\
01-04-14  & 56749.5 & L &         -       & $1.81\pm0.01$ &$ -0.057    \pm 0.023$  & $2.31\pm0.02$    &  $7.1\pm0.1$      & $23.3\pm0.4$   \\

\tableline                                  
\end{tabular}}
\tablecomments{$^a$ A - Copernicus telescope, Asiago; L - Cassini telescope, Loiano.}
\end{center}
\end{table*}

 The uncertainties on line  fluxes  are estimated by  computing the flux difference $\delta = |f^{(2)}-f^{(1)}|$  between the two exposures taken at the  same  epoch.  Since the exposure time is roughly constant at all epochs,  and fluxes  are computed as the average between two exposures, we adopt as uncertainty  on the flux $\sigma_f=\langle\delta^2\rangle^{1/2}/2$, where the angular brackets indicate the average over the entire set of measurements.
The fractional values of $\sigma_f$ for continuum,  C{\scriptsize III]}, C{\scriptsize IV}  fluxes are  0.008, 0.021 and 0.015  respectively.
Direct photometry of the field was also obtained at most epochs, to get an independent measure  of the quasar luminosity changes, relative to the reference star,  and to check the stability of the reference star against other objects in the field. Typically, at each epoch, R band photometry was obtained with exposures of 400 s at Asiago Observatory.  Photometry is also available in R and/or V bands, from Loiano Observatory.

From the average spectrum we can estimate that the contribution  of  C{\scriptsize III]} line to the R and V  magnitude is about  0.04 mag and 0.03 mag respectively,  while the contribution of C{\scriptsize IV}  is negligible in both bands. Thus we can use photometric data to measure continuum variations  without altering significantly the continuum-line cross-correlation function. 
In doing this, we convert V  to R' magnitudes assuming R'(V)$\equiv$ V-$\langle$V-R$\rangle$, where the angular brackets indicate the average over those epochs when both R and V are measured. This conversion assumes that the V-R quasar color is constant. We have checked that r.m.s. color changes are  of the order of 0.007 mag. Such color changes could slightly affect the amplitude of the continuum-line cross-correlation only in the case of C{\scriptsize III]},  but cannot affect significantly the estimate of the time delay. A measure of the continuum changes can be obtained also from the spectra. This has been done by fitting with a single straight line,  $\log F^{cont}(\lambda)=-a~ \log \lambda+b$, the four data points which define the local continua (shown in figure \ref{fig:spec}). From these fits a conventional spectral continuum, at the peak wavelength of the \citet{Bess90} V band,  $F_C\equiv F^{cont}_{\lambda}(5300 {\rm\AA})$ has been defined. For the subsequent analysis (see Sect. 3), continuum flux changes have been referred to the epoch $t_{ref}$ (MJD=53047.5) and expressed as magnitude changes $\delta m_C(t_i)= \-2.5  \log [F_C(t_i)/F_C(t_{ref})]$. Similarly for line fluxes. Continuum changes obtained from broad band photometry have been reduced to the same scale defining $\delta R(t_i)=\Delta R(t_i)+\langle \delta m_C - \Delta R \rangle$, where the average is taken over all the epochs when both photometric and spectroscopic data  are available.

Table \ref{tab-journal} reports the results. Column 1 : date; column 2: modified julian date MJD; column  3:  Telescope;  columns  4 and 5:  V and R band magnitude differences with respect to the reference star; column 6: continuum changes $\delta R$ obtained from broad band photometry; column 7: continuum specific flux $F_C$ at $\lambda=5300 {\rm\AA}$; columns 8 and 9:  C{\scriptsize III} and C{\scriptsize IV]} line fluxes respectively.

\begin{figure}
\centering
\includegraphics[scale=0.44]{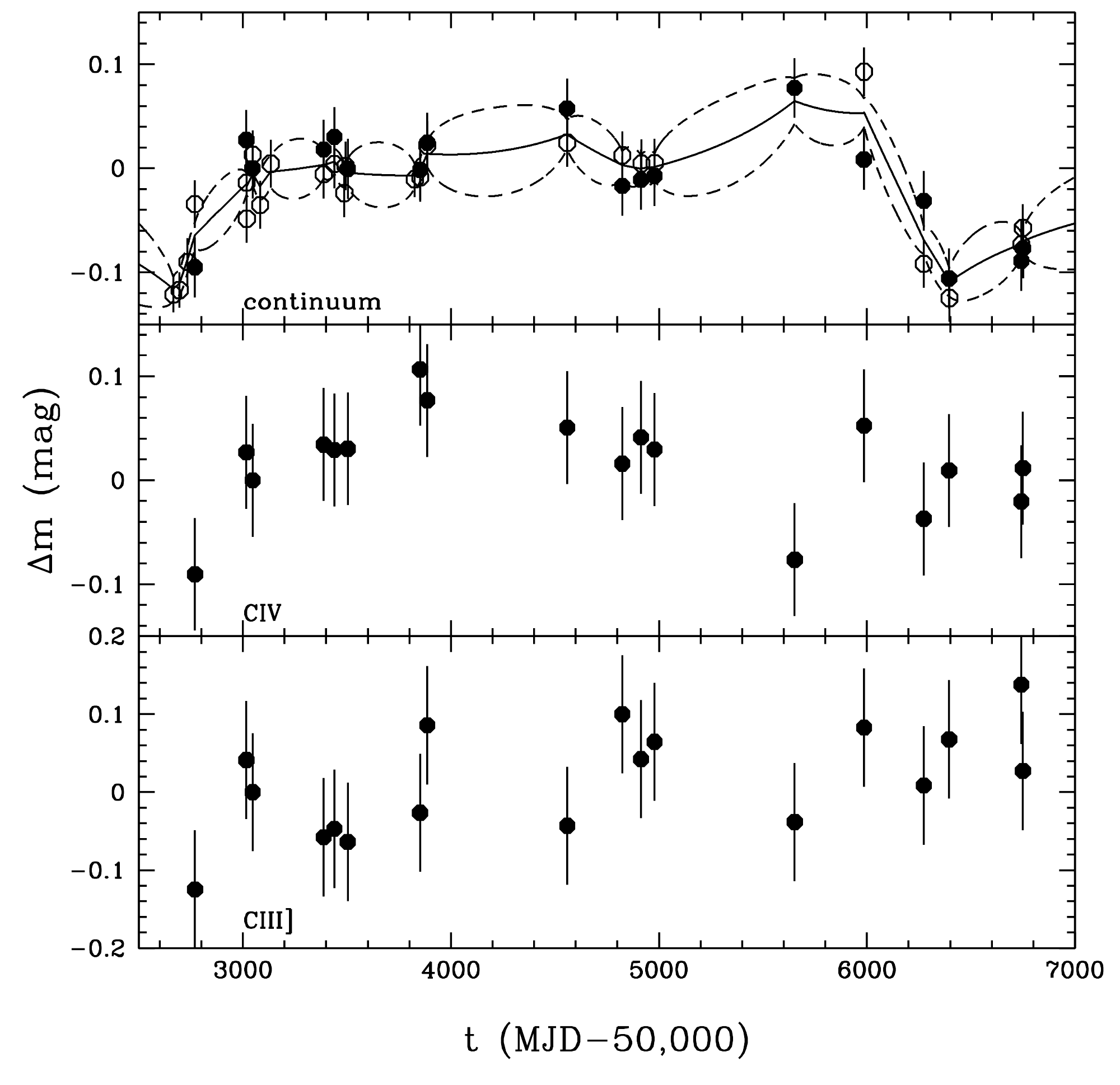}
\caption{Magnitude changes as a function of time.  {\it Upper panel}: continuum changes from spectrophotometry $\delta m_C$ (filled circles) and from broad band photometry $\delta R$ (open circles).   Data interpolation (continuous line) and the relevant 1-$\sigma$ uncertainty (dashed lines) according to the method of \citet{Zu11}. {\it Middle panel}: C{\scriptsize IV}(1549{\rm\AA})~ line. {\it Bottom panel}: C{\scriptsize III]}(1909{\rm\AA})~line.}
\label{fig:curve}
\end{figure}

Figure \ref{fig:curve} reports the light curves in magnitude for continuum  and emission lines.

\section{Measuring the reverberation time lags}

The time lag $\tau_l$ of the emission line variation with respect to  continuum changes  is measured by  the centroid of the continuum-emission line cross-correlation function, 
which can be computed through the discrete correlation function (DCF)  \citet{Edel88} or by  interpolating the light curves \citep{Gask87,Whit94}. Both methods provide consistent results for  well-sampled light curves. In the case of poor sampling, both methods present technical problems  \citep[see the review by][]{Pete93}.
In particular the DCF becomes less sensitive to real correlations. Moreover, the estimate of a confidence interval on the measured time lag, which can be obtained by the z-transform method developed by \citet{Alex97}, requires to eliminate from each DCF bin all points corresponding to pairs of epochs having a measure in common.
This causes the loss of part of the information contained in the data, so that the method may be not applicable if the total number of observations is too small. 

We adopt a  methodology, called Stochastic Process Estimation for AGN Reverberation (SPEAR), developed by  \citet {Zu11}. The statistical basis of the method was introduced by \citet{Pres92} and \citet{Rybi92}. A subsequent modification and application to RM of NGC~5548 is discussed in \citet{Rybi94}. For  a detailed description of the SPEAR we refer to the paper of \citet{Zu11}. An upgraded version called JAVELIN, which allows photometric RM, is discussed in \citet[][b]{Zu13b}. Here we recall a few important points which motivate our choice to adopt  this method  to measure reverberation time lags in our case. First of all,  it makes use of interpolation, which is essential for us, given the small number of data points. But, while a simple linear interpolation is based on two nearby points, here  the entire dataset contributes  to each interpolated point, through weights which are statistically determined from the correlation functions of the data. Moreover, statistical uncertainties are assigned to each interpolated value. The uncertainties tend to the  measurement errors in correspondence with the data points, and become larger and larger when the distance from the neighbor data points increases (see Figure \ref{fig:curve}). The main assumption of the method is that the emission-line flux variations $l(t)$ are scaled, smoothed, and time-shifted  versions of the continuum  variations $c(t)$, obtained through a transfer function $\Psi(t)$:
\begin{equation}
l(t)=\int dt' \Psi(t')c(t-t'),
\end{equation}
In our analysis, we assume for $\Psi(t)$  the simple form adopted by \citet{Zu11}:
\begin{equation}
\Psi(t)=A/\Delta, ~~|t-\tau_l| \leqslant \Delta; ~~\Psi(t)=0~~ {\rm elsewhere},
\end{equation}
where $A$, $\tau_l$ and $\Delta$ represent the attenuation, the line-continuum lag and the temporal width respectively. A maximum likelihood code determines the attenuation, smoothing and time lag parameters. The resulting emission line delay $\tau_l$ does not depend strongly  on the form assumed for $\Psi(t)$ \citep{Rybi94}. The correlation functions of the data are represented by parametric models whose parameters are also determined by likelihood maximisation.  This allows adding information deduced from existing data on the statistical properties of QSO light curves. In fact, it has been shown that a damped random walk (DRW)  process is a good representation of QSO variability \citep{Kell09,Kozo10,MacL10,Zu13a}.
The DRW auto-correlation function of the continuum changes $c(t)$ has the form:
\begin{equation}
\langle c(t) c(t+\tau_l) \rangle=\sigma^2 exp(-|\tau_l|/\tau_d),
\end{equation}
where $\tau_l$ is a time lag, $\tau_d$ is the damping time scale, $\sigma$ is the variability amplitude, and angular brackets indicate the ensemble average. 
Another important feature of the SPEAR method is that the light curves of more lines can be included in the same fitting procedure. This provides more stringent constraints which allows  better choice among the local likelihood maxima in the  space of parameters.

A single fitting procedure determines  the eight parameters:
$\sigma$ and $\tau_d$ for the continuum,  the time  lags $\tau_{CIV}$, $\tau_{CIII]}$, the amplitudes $A_{CIV}$, $A_{CIII]}$ and smoothing parameters  $\Delta_{CIV}$, $\Delta_{CIII]}$, for the two lines respectively. 
\begin{figure}
\centering
\includegraphics[scale=0.44]{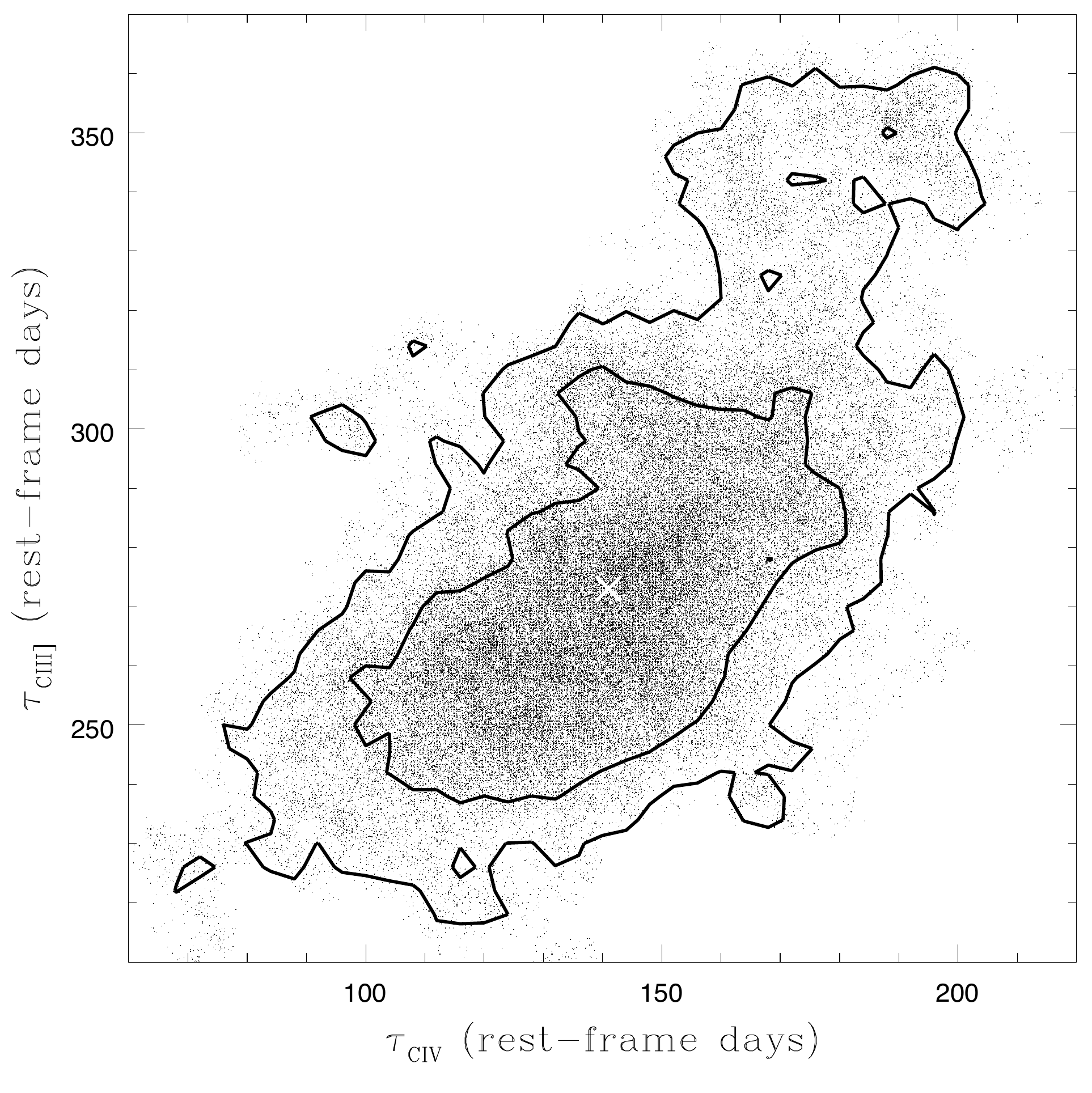}
\caption{The distribution of points generated by 10$^5$ MCMC iterations in the ($\tau_{CIV}$,$\tau_{CIII]}$) plane. Contours correspond to 68\% and 95\% confidence levels. The white cross indicates the median values of the marginal distributions of the two parameters.
}
\label{fig:like-map}
\end{figure}
SPEAR adopts a Bayesian method to determine the confidence interval in the parameters space. Once the values of the parameters which maximize the likelihood are determined, random increments obtained from  {\it prior} statistical distributions are applied to all parameters and the likelihood is re-evaluated. Following a Markov Chain Monte Carlo (MCMC) method \citep[see][and refs therein]{Pres07}, the process is iterated and a {\it posterior} distribution of acceptable parameters is produced. Figure \ref{fig:like-map} shows the distribution of points in the ($\tau_{CIV}$,$\tau_{CIII]}$) plane after $10^5$ MCMC iterations, and the corresponding 68\% and 95\% confidence regions. 
Figure \ref{fig:histo} shows the relevant posterior distributions obtained separately for  $\tau_{CIV}$ and $\tau_{CIII]}$. From these we take as fiducial estimate of the fitting parameters their median values, with the uncertainties defined by the 68\% confidence intervals: $\tau_{CIV}=142\pm^{26}_{25} $ days,  $\tau_{CIII]}=273\pm^{30}_{21}$  days in the QSO rest-frame.

\begin{figure}
\centering
\includegraphics[scale=0.44]{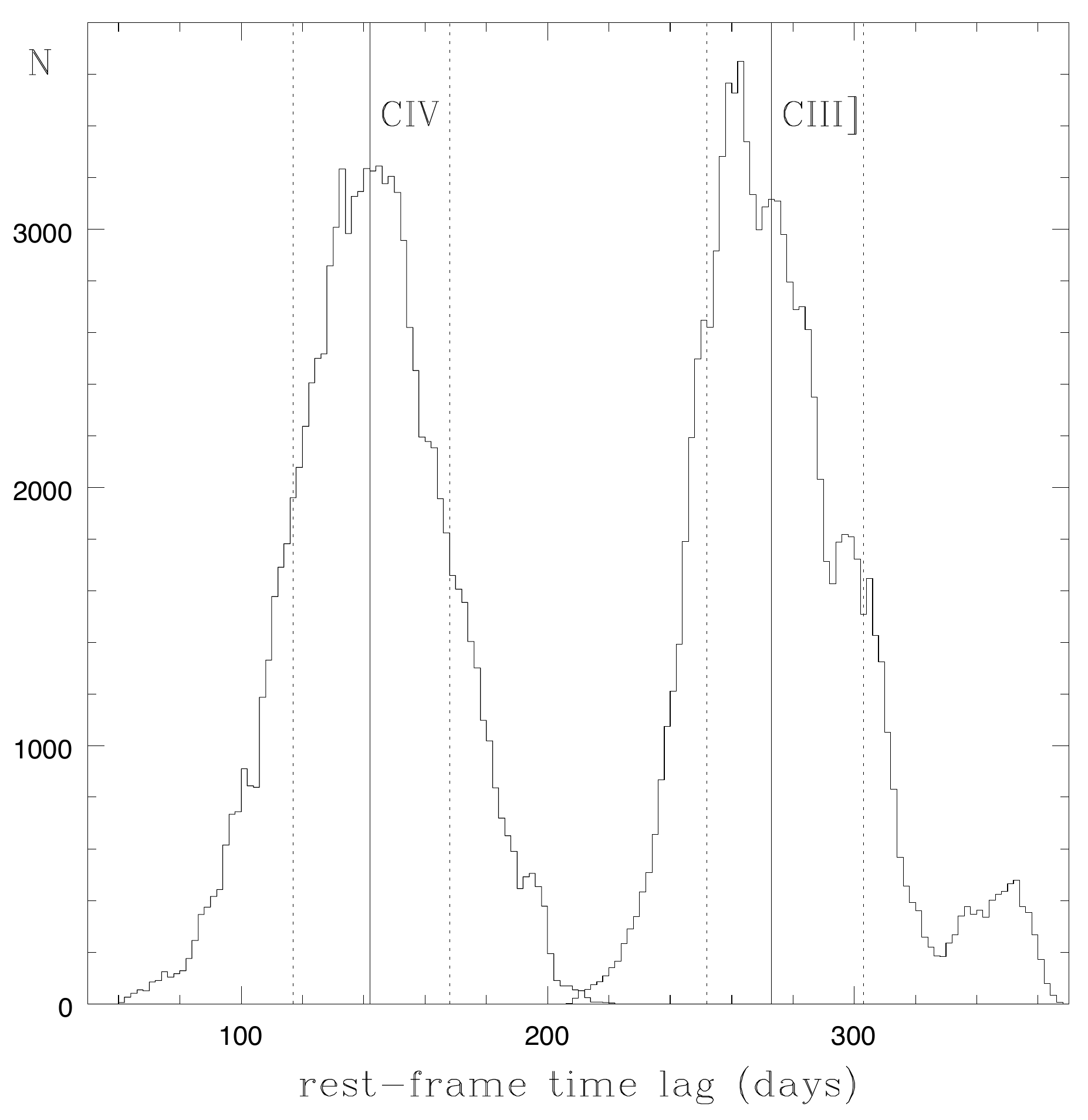}
\caption{Posterior distributions of the rest-frame time lags  $\tau_{CIV}$ and $\tau_{C_III]}$  produced by the SPEAR method with $10^5$  MCMC iterations. Continuum  vertical lines indicate the median values and the dashed lines indicate the 68\% confidence intervals.
}
\label{fig:histo}
\end{figure}

With respect to our  preliminary results \citep{Pern14}, the present analysis differs because: i)  we fit simultaneously both C{\scriptsize IV} and C{\scriptsize III]} light curves  ii) the continuum light curve includes  the available V and  R band photometry, together with the continuum variations measured from the spectra, and iii) we have included photometric and spectroscopic data of three most recent epochs.
Considering the small number of points and the uneven sampling, with two main gaps for  MJD (53900,54500) and MJD (55000,55600), it is worth wondering whether the likelihood maxima are real or are determined by the sampling pattern. To this end, we performed a Monte Carlo simulation, generating $N=1000$   mock, uncorrelated, continuum and emission line light curves,   assuming the same set of sampling epochs, r.m.s. variability amplitude, and measurement uncertainties. We applied the SPEAR procedure to the $k$-th set of light curves and,  once  $\sigma$ and $\tau_d$ were fitted,  we produced a likelihood ``image"  $\mathcal{L}_k(\tau_l,\Delta_l)$  as a function of the lag $\tau_l$ and the smoothing parameter $\Delta_l$, ($l=CIV,CIII]$) .
Then we analyzed the sum $\mathcal{L}\equiv\Sigma_{k,1}^N\mathcal{L}_k$, which must show local maxima 
in correspondence of the points ($\tau_l$,$\Delta_l$) where maxima occur  more frequently in the  simulations, thus indicating the effect of  the fixed sampling pattern.
The result shows that  local maxima do exist but, with respect to the case of real data: i) they are confined to much lower values of $\Delta_{CIV}$ and $\Delta_{CIII]}$,  ii)  they are less pronounced  and, most important, iii) they are not located in the same position of the ($\tau_{CIV}$-$\tau_{CIII]}$) plane where they occur in the case of real data.
 
We can conclude that it is very unlikely that  the local maxima related with the sampling pattern may produce the maxima obtained in the case of the measured light curves.
Thus we  assume this result as first evidence that  the values of  $\tau_{C~IV}$ and $\tau_{CIII]}$  for PG~1247+267 are due to real echo lags. 

\begin{figure}
\centering
\includegraphics[scale=0.44]{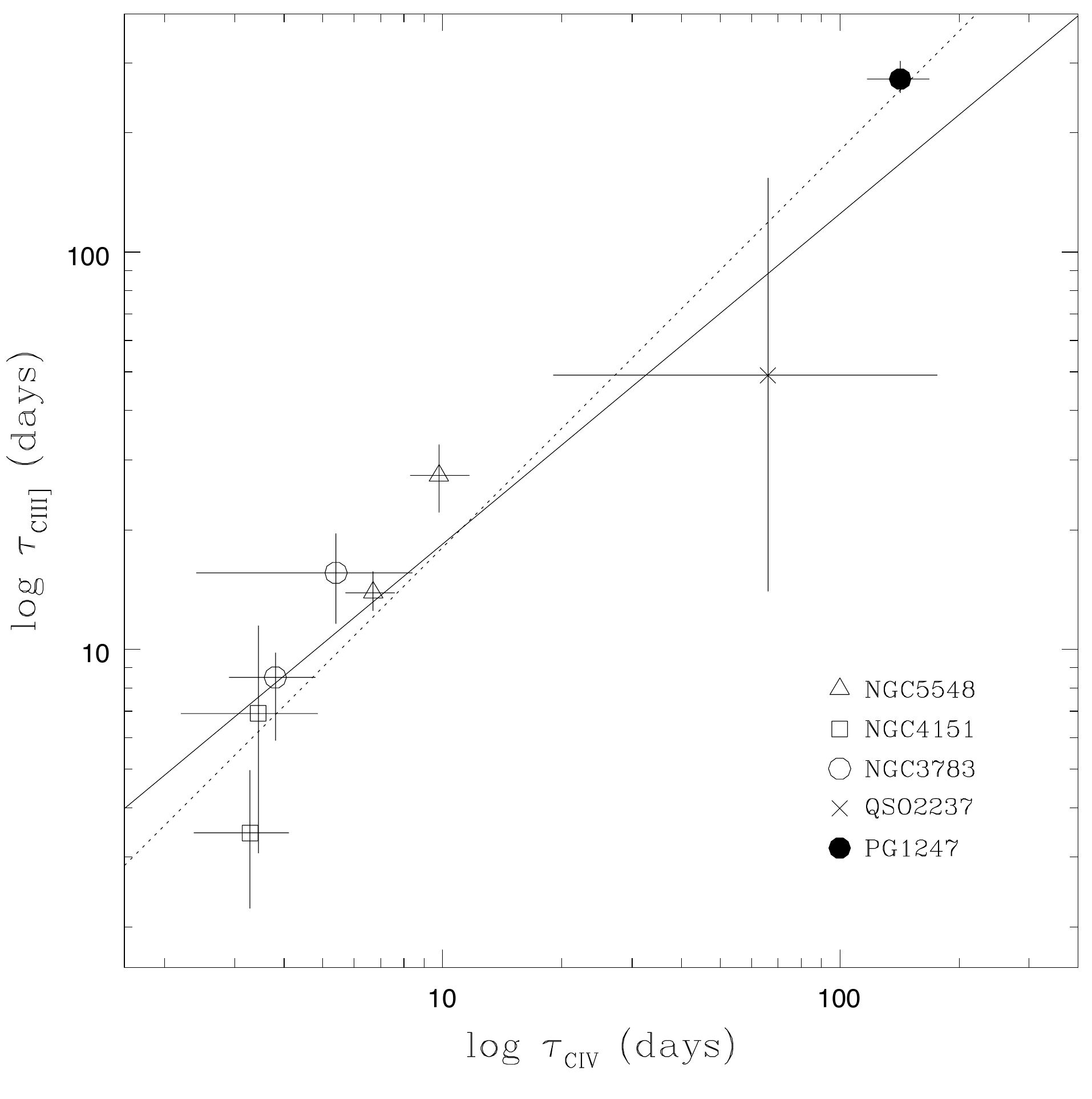}
\caption{ Reverberation time lags $\tau_{C III}$ versus $\tau_{C IV}$  for NGC~5548, NGC~3783, NGC~4151, PG~1247+267 (our estimate). The emission region sizes \citep{Slus11}, converted to time lags, are also reported for the quasar QSO 2237+0305. Straight lines represent linear fits with errors on both variables: continuous line with two free parameters, dotted line with fixed unitary slope. 
}
\label{fig:tau_civ_ciii}
\end{figure}

We can compare them with the few corresponding RM results available in the literature. Measures of both  C{\scriptsize IV} and C{\scriptsize III]}  time lags from RM exist  for 3 Seyfert nuclei:  NGC~5548 \citep{Pete99}, NGC~3783 \citep{Onke02}, NGC~4151 \citep[][and refs therein]{Metz06}, all less luminous than $\lambda L_{\lambda}(1350 {\rm\AA}~) \approx~4~\times~10^{43}$~erg ~s$^{-1}$. On average, the ratio of the time lags of these two lines is $\tau_{C III}/\tau_{C IV}\approx 2$. While there is no reason to expect that a QSO,  $10^4$ times more luminous,  should show approximately the same ratio, it is interesting to note that in the case of PG~1247+267 we obtain $\tau_{C IV}/\tau_{C III]}\sim 2$, i.e. the typical distance from the continuum source of the  emission region of the semi-forbidden C{\scriptsize III]} line   is about twice the distance of the  C{\scriptsize IV} emission region.
  The situation is summarized in Figure \ref{fig:tau_civ_ciii}, where we report also the relevant time lags for the QSO 2237+0305, as deduced from \citet{Slus11}, who estimate the size of the C{\scriptsize IV} and C{\scriptsize III]} emission regions on the basis of microlensing. The corresponding point looks roughly consistent with the general trend, despite the relevant sizes are of the order of 100 times larger than those of Seyfert galaxies. A straight line fit to the data points in Figure \ref{fig:tau_civ_ciii},  $\log \tau_{CIII]}=a \log \tau_{CIV} + b$ gives $a=0.83\pm 0.21$ and $b=0.43\pm 0.19$. A fit with fixed unitary slope gives $\tau_{CIII]}/ \tau_{CIV}=1.8\pm 0.5$.  The very existence of  this relation can be taken as a second suggestion that we are probably measuring real reverberation time lags. 
This result would mean  that the ionization stratification in Seyfert nuclei and luminous quasars is similar.

\begin{figure}
\centering
\includegraphics[scale=0.44]{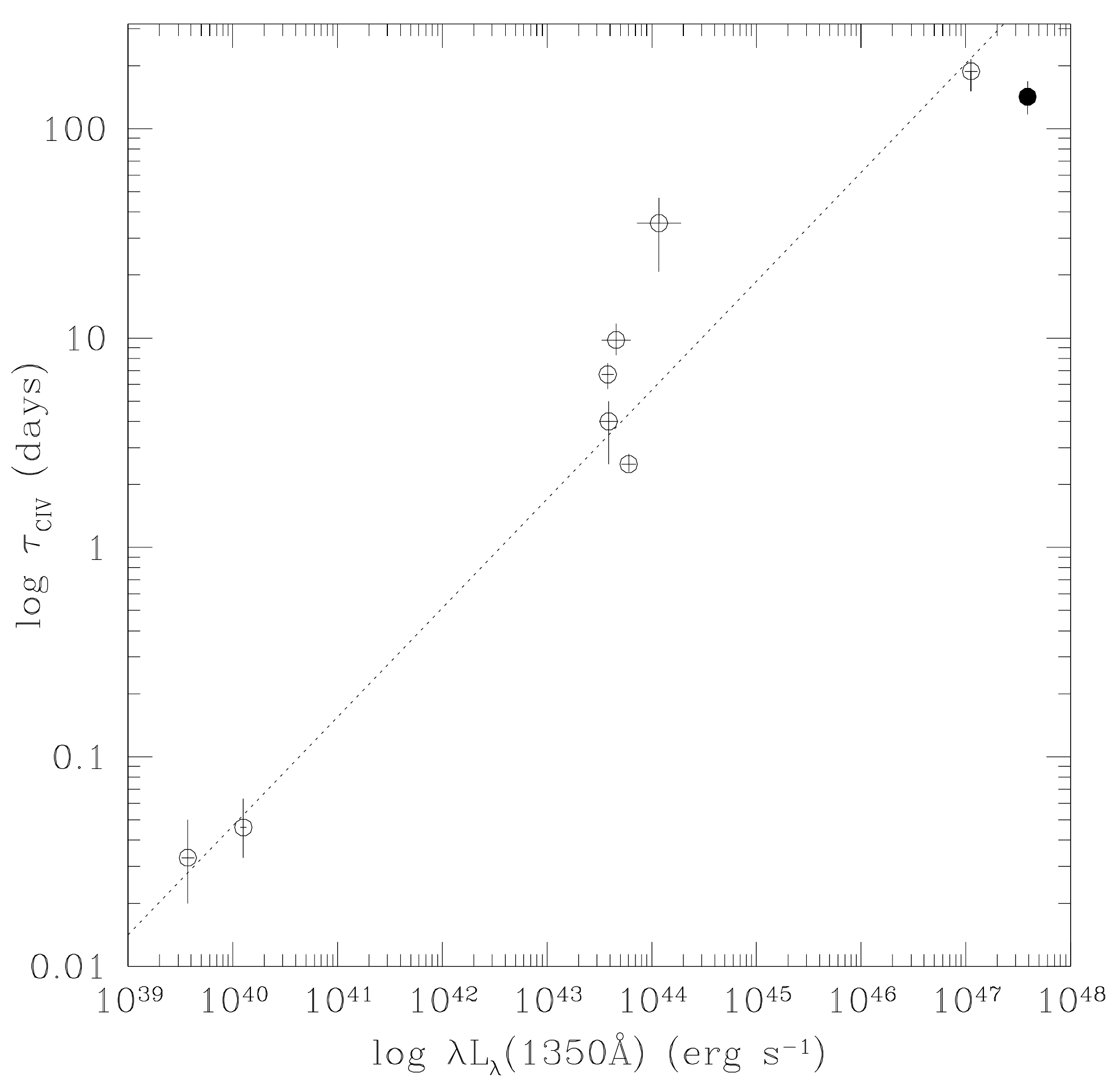}
\caption{Size-luminosity relation obtained from C{\scriptsize IV} emission line and UV continuum. {\it Open circles}: data from \citet{Pete05,Pete06} plus the values for S5 0836+71 from \citet{Kasp07};
{\it filled circle}:  our result for PG~1247+267. The dotted line represents the linear fit  obtained by \citet{Kasp07} with the FITEXY method \citep{Pres92}.
}  
\label{fig:size-lumin}
\end{figure}

Figure \ref{fig:size-lumin} shows the relation between the emission radii  $R_{CIV}=c \tau_{CIV}$ 
and the luminosity $\lambda L_{\lambda}(1350{\rm\AA})$ for all the objects studied  so far. For their second brightest object S5~0836+071, \citet{Kasp07} give $\lambda L_{\lambda}(1350 {\rm\AA}) = 1.12  \pm 0.16 \times 10^{47}$ erg s$^{-1}$ and a tentative value of the  C{\scriptsize IV} emission line delay $\tau_{CIV} = 188 \pm^{27}_{37}$ days in the quasar rest frame and discuss the slope of the $\tau_{CIV}-\lambda L_{\lambda}$  relation obtained by different fitting procedures. The fit, shown in Figure \ref{fig:size-lumin},  they obtain with the FITEXY algorithm \citep{Pres92}, corresponds to a slope  $\gamma=0.52\pm 0.04$.

PG~1247+267 is the brightest QSO ever analyzed for reverberation and has  $\lambda L_{\lambda}(1350{\rm\AA})  = 3.92 \pm 0.02\times 10^{47}$ erg s$^{-1}$, 
deduced from \citet{Shen11}, thus it is   3.5 times more luminous than S5~0836+071.  Compared with the extrapolation of the lag-luminosity relation in Figure \ref{fig:size-lumin}, PG~1247+267 should show a reverberation lag of  about 400 days, i.e. $\sim$ 3 times larger than observed.
If confirmed, this would imply a  decrease of about 10\% of the slope $\gamma$ of the $\tau_{CIV}-\lambda L_{\lambda}$ relation, whose significance is marginal, however,  due to the  relatively large dispersion of the still small number of points in the lag-luminosity diagram for C{\scriptsize IV}.
In the case of S5~0836+71, \citet{Kasp07} obtain  $\tau_{CIII]}$ consistent with zero. A possible explanation may be that   they do not use an interpolation of the light curves, which becomes necessary when the total number of points is small.

\section{Estimating the virial mass}

\begin{table*}[t]
\caption{Reverberation results for PG~1247+267}            
\label{tab-reverb}      
\centering                          
\begin{tabular}{l l l l l l}        
\hline\hline                
\rule{0pt}{2ex}
Emission line& $\tau_l^{~~a}$&FWHM$_{mean}$&$\sigma_{l,mean}$&FWHM$_{rms}$&$\sigma_{l,rms}$ \\ 
 & & \multicolumn{4}{c} \hrulefill  \\
& days & \multicolumn{4}{c} {km s$^{-1}$}  \\

\hline                        
\rule{0pt}{2ex}
C{\scriptsize IV}\, $\lambda$1549 &142 $^{+26}_{-25}$ &4939 $\pm$ 117 &2673 $\pm$ 20 &4568 $\pm$ 1338 & 2104 $\pm$ 540\\
\rule{0pt}{3ex}
C{\scriptsize III]}\, $\lambda$1909 &273 $^{+30}_{-21}$ &5224 $\pm$  ~~63 &2365 $\pm$  ~15 &  4752 $\pm$ ~1156& 1899 $\pm$ 713  \B \\

 \hline                                  
\end{tabular}
\tablecomments{$^a$ In the rest-frame}
\end{table*}

From the emission radii  $R_{CIV}=c \tau_{CIV}$ and $R_{CIII]}=c \tau_{CIII]}$, we can try and measure the BH mass through the virial relation. Unfortunately, the use  of C{\scriptsize IV} emission line for mass estimation appears problematic \citep{Netz07}, since the  profile of this line reveals the contribution  of different components, whose relative weight vary so much from object to object, that the shape of this line turns out to be a good indicator of different AGN types \citep{Sule07}.
In principle the  velocity $\Delta V$ appearing in the virial relation $M_{BH}=f R \Delta V^2/G$ could be identified with the FWHM of the emission line, or with the r.m.s. velocity dispersion along the line of sight $\sigma_l$ \citep{Pete04}.

Under the assumption of isotropic velocity field, the kinetic energy is $K=3/2 M \sigma_l^2$, but the numerical factor may  differ from 3/2 depending on anisotropies, possibly related to the shape of the broad line clouds system.
As a consequence, the numerical factor can be different for different emission lines, even in the same AGN.  In the case of a Gaussian line profile,  FWHM/$\sigma_l$=2.35, but may be different for different profiles. All the numerical factors are absorbed in the factor $f$ of the virial relation and contribute to the uncertainty of mass determination,  both in the case of RM and SE measurements. 
In RM experiments, where several spectra taken at different times are available, it is possible to compute the r.m.s. spectrum \citep{Pete98}.

The non-variable parts of the spectrum, or those which vary on time-scales much longer than the experiment, do not contribute
to the r.m.s. spectrum. The r.m.s. spectrum of PG~1247+267 is shown in the lower panel of  Figure \ref{fig:spec}.  The uncertainty on $\sigma_l$ can be computed by applying a bootstrap procedure described in \citet{Pete04}. The reason to use the r.m.s. instead of the average spectra  is to avoid : i) the underestimates of $\Delta V$ caused by narrow emission line component, and  ii) the effect of non virial outflows, which are expected to vary on time scales longer than reverberation time lags \citep[see][]{Denn12}. 

Our results are summarized in Table \ref{tab-reverb}, where the reverberation time lags  for both C{\scriptsize IV} and C{\scriptsize III]}  are reported,  together with the relevant FWHM and  $\sigma_l$, computed from mean and r.m.s. spectra. The virial products $\tau_{CIV} \Delta V_ {CIV}^2$ and $\tau_{CIII]} \Delta V_{CIII]}^2$ appear consistent with the same black hole mass. We assume this fact as a third evidence that we are measuring real reverberation lags.
 
The numerical factor $f$ in the virial relation can be determined empirically, by calibrating the RM masses through the $M_{BH}$-$\sigma_*$ relation \citep{Ferr00}, as done  for H$\beta$ RM by \citet{Onke04} who finds $f$=5.5. 
From the calibrated H$\beta$ RM masses,  \citet{Vest06} computed statistical scaling relations permitting SE mass determinations  on the basis of the luminosity $L_{\lambda}(5100 {\rm\AA})$ and H$\beta$ line width. Scaling relations for SE mass determination based on $L_{\lambda}(1350 {\rm\AA})$ and C{\scriptsize IV} line widths, normalized to H$\beta$ masses, are also provided.
However, as mentioned above, the C{\scriptsize IV} line shape varies from object to object, and this causes a large spread in the measured masses around the scaling relation.
A more accurate relation can be obtained by analysing the line shape parameter $S=FWHM/\sigma_l$, as computed both from the average spectrum and from the r.m.s. spectrum $\sigma(\lambda)$: $S_{mean}$ and $S_{rms}$ respectively. \citet{Denn12} pointed out that, while the ratio $S_{mean}/S_{rms}$ is of  order one for  the H$\beta$ line,  it is different for different values of $S_{mean}$ in the case of C {\scriptsize IV} (see her Figure 2). This is interpreted in terms of outflowing, non reverberating component not contributing to the r.m.s. spectrum.  A correction, dependent on $S_{mean}$, is proposed to reduce the masses derived from C{\scriptsize IV} to those derived from H$\beta$ (Eq.1 in \citet{Denn12}): 
$\log M^{corr}_{CIV}=\log M^{orig}_{CIV}+0.219-1.63 \log (FWHM_{CIV}/\sigma_{CIV})$.

In the case of PG~1247+267,  $S_{mean}$= 1.85 and the value $S_{rms}/S_{mean}=2.2/1.85=1.18$ is close to one, meaning that the use of the r.m.s. spectrum does not change much the value of $S$, namely the effect of a non reverberating component, though present,  is small. This quantifies the similarity of C{\scriptsize IV} emission line in the average and r.m.s. spectra, already noticed in Figure \ref{fig:spec}.  It is interesting to note that S5~0836+071 has $S_{mean}=1.94$ (instead of 1.85), indicating  that the shapes of the C{\scriptsize IV}  line of the two objects are similar. This suggests  that  the relevant virial factors $f$ too are similar,   so that  the ratio of the masses  is approximately the ratio of the virial products  $\tau_l\Delta V^2$, whatever the definition of $\Delta V$.

\begin{table}[b]
\caption{Mass estimates for PG~1247+267}            
\label{tab-mass}      
\centering                          
\begin{tabular}{l l l}        
\hline\hline                
\rule{0pt}{2ex}
Emission line&  $M^{mode}_{BH}$& $M^{cent}_{BH}$ \\
 & \multicolumn{2}{c} \hrulefill  \\
& \multicolumn{2}{c} {10$^8 M_{\odot}$} \\

\hline                        
\rule{0pt}{2ex}
C{\scriptsize IV}\, $\lambda$1549 & 6.7$^{+5.0}_{-1.1}$   &  8.3 $^{{\bf+3.4}}_{{\bf-2.7}} $ \\
\rule{0pt}{3ex}
C{\scriptsize III]}\, $\lambda$1909 & 10.5$^{+17.2}_{-9.1}$   &9.9 $^{+17.8}_{-8.5}$ \B \\

 \hline                                  
\end{tabular}
\end{table}
 
For the luminous quasar S5~0836+071, with $\lambda L_{\lambda}(1350 {\rm\AA})= 1.12 \pm 0.16 \times 10^{47}$ erg  s$^{-1}$,  \citet{Kasp07}  obtained $\tau_{CIV}=188\pm27$ days in the rest-frame. They define  $\Delta V$ as the FWHM$_{mean}$ and, using  $f=3/4$ in Eq.2 of \citet{Kasp00}, they obtain a mass $M_{BH} \sim 2.6 \times 10^9 M_{\odot}$, which corresponds to a virial product  $c \tau_{CIV} \Delta V_ {CIV}^2\sim 3.5 \times 10^9 M_{\odot}$. Adopting the same definition of $\Delta V$,   we obtain for PG~1247+267 $M_{BH}\sim 6.7  \times 10^8 M_{\odot}$, i.e. about 5 times smaller, in spite of its higher luminosity.
 
Using as definition of $\Delta V$ the value of $\sigma_l$  obtained from the r.m.s spectra, \citep{Onke04}  obtained an average virial factor $f=5.5$. More recent estimates  \citep[see][and refs. therein]{Panc13} provide different values, but we adopt conventionally the more commonly used $f=5.5$ for the subsequent comparison with the literature. Based on this calibration, and using $\sigma_l$  obtained from the r.m.s spectrum (see Table \ref{tab-reverb}) we obtain $M_{BH}\sim 6.7 \times 10^8 M_{\odot}$.
 
We stress that, being the distribution of the  $\Delta V^2$ and $\tau_l$ asymmetric, some care is needed in deriving the fiducial mass values and the relevant confidence intervals.  For this purpose, we computed a  probability distribution of the virial product, as a function of   $\Delta V^2$ and $\tau_l$,  by multiplying the posterior distribution of  $\tau_l$ (see Figure 4) and the statistical distribution of $\Delta V^2$ obtained by the bootstrap method. From this we derived a posterior distribution of the black hole mass.  In addition to the modal mass value $M_{BH}^{mode}$, we report in Table \ref{tab-mass} the value computed as the centroid of the posterior distribution, together with the asymmetric errors at 68\% confidence level.
Hereinafter we will use as our best estimate of the virial mass $M_{BH}^{cent}=8.3 ^{+3.4}_{-2.7} \times 10^{8} M_{\odot}$. The corresponding mass values obtained from C{\scriptsize~III]} are also reported in Table \ref{tab-mass}.

We  can  compare our result  with the summary of  BH masses, known from RM, versus $\lambda L_{\lambda}(1350{\rm\AA})$, shown in Figure 5 of \citet{Chel12}. From this comparison it appears that the new point, we are adding at the highest luminosity, corresponds to a mass which is roughly a factor 20 smaller with respect to the extrapolation of the general trend, which would predict a mass of the order of $2 \times 10^{10}$M$_{\odot}$. 
The scatter of points around the M$_{BH}$-$\lambda L_{\lambda}(1350{\rm\AA})$ relation is partly intrinsic, due to the fact that different AGNs may be emitting at different Eddington ratios, and partly caused by the uncertainty on the  $f$ factor appropriate for individual objects. Thus a deviation from the average scaling relation of a factor $\sim$ 20 is not surprising. However it deserves further discussion.

\section{Discussion and summary}

The number of spectral observations is still small  and  requires caution in deriving any conclusion. However,  three independent circumstances suggest that we are probably  measuring real reverberation time lags: i) despite it is reasonable to expect that the likelihood maxima might be determined by the uneven temporal sampling, Montecarlo simulations with mock random light curves, and the same sampling pattern, do not produce  likelihood maxima in the  same region of the parameters space; ii) the measured C{\scriptsize IV} and  C {\scriptsize III]} reverberation time lag appears consistent with the $\tau_{C IV}$-$\tau_{C III]}$ relation derived from the data available in the literature; iii) the virial products for C{\scriptsize IV} and C{\scriptsize III]} lines appear consistent with the same black hole mass. 

Thus, assuming that  the measured  $\tau_{CIII]}$ and $ \tau_{CIV}$ are real, we can derive some tentative conclusions.

The fact that the approximate relation $\tau_{CIII]} \sim 2 \tau_{CIV}$ (see Figure \ref{fig:tau_civ_ciii}) extends from  objects with luminosity $ \lambda L_{\lambda}(1350 {\rm\AA})$
from $\approx  4 \times 10^{39}$ to $\approx 4  \times 10^{47}$erg s$^{-1}$, if confirmed, would be a first direct evidence that the ionization stratification in luminous QSOs is similar to that found in Seyfert galaxies.

The relatively small  $\tau_{CIV}$, about  0.3 of the value expected from the extrapolation of the $\tau_l-L$ relation, tends to produce a small virial mass. The problem is made more severe by the small value of the line widths,  roughly 2/3 of that of S5~0836+071 which is  3.5 times less luminous. This  appears clearly when we compute the Eddington ratio $L_{bol}/L_{Edd}$, which contains a further uncertainty  deriving from the estimate of the bolometric correction. \citet{Kasp07} adopt the bolometric correction of \citet{Marc04} (Eq.21) which refers to the luminosity $\nu_B L_{\nu B}$. For PG~1247+267 we obtain 
$\nu_B L_{\nu B}=2.0 \times 10^{47}$erg s$^{-1}$, interpolating between the values of $\lambda L_{\lambda}(2500 {\rm\AA})$ and $\lambda L_{\lambda}(5100 {\rm\AA})$ provided by \citet{Kraw13}.  
The resulting  correction is  $L_{bol}/\nu_B L_{\nu_B}=5.28$, leading to  $L_{bol}=1.06 \times 10^{48}$ erg s$^{-1}$  and  $L_{bol}/L_{Edd} = 9.8 $ (after a small correction for the different cosmology we adopt). 
A similar result,  $L_{bol}/L_{Edd} \sim 10.4$  is found adopting the bolometric correction $L(1\mu-2keV)/\nu L_{\nu}(2500 {\rm\AA})= 3.5$  with $\log \nu L_{\nu}(2500 {\rm\AA})=47.50$ from \citet{Kraw13}. 

We stress that  this large Eddington ratio is due only in part to the small size of the BLR. In fact, even the SE mass estimate, which is independent of reverberation lag, produces for PG~1247+267 an Eddington ratio in the range 1.2 - 3, depending on the use of the line shape correction \citep{Denn12} and the different choices of bolometric correction.

As discussed in the previous section,  a possible  origin of a too small line width may be the presence of a narrow emission line component, or the contribution  of a possibly non variable and non virial wind component. This suggested to use, as we did, the r.m.s. spectrum. Besides this, orbits of BLR clouds are unlikely to be oriented randomly,  as suggested by various evidences reviewed by \citet{Gask09}, and sources viewed at a low  inclination  angle (nearly face-on) show  a small  FWHM, leading to a systematic underestimation of the black hole mass by  a factor up to $\sim$10 \citep{Marz12}. 

For 5 Seyfert nuclei, \citet{Panc13} compute line profile models,  which depend  on the opening angle of the cloud distribution,   for different values of the inclination angle with respect  to the axis of the accretion disk. The relevant virial factor $f$,  to be applied when using $\sigma_l$ from r.m.s. spectra, can be as high as 50 for an inclination angle of 8 degrees. 
Thus, a plausible high value of the virial factor $f$ could easily bring the Eddington ratio towards more common values, without, however, explaining the relatively small  reverberation time lag.

The effects of orientation on the characteristics of the C{\scriptsize IV} line have been investigated by \citet{Runn14}. By comparison with their Figure 3,  the small amplitude  relatively to Si~{\scriptsize IV} and narrow line width found in the spectrum of PG~1247+267  suggest, in fact, a small inclination  angle, supporting a high $f$ value.  A more quantitative evidence would require, however, a dynamical modeling of the type presented by \citet{Panc13}, and a comparison with velocity resolved RM, not feasible with our present data.

A different, and apparently trivial,  explanation of the high Eddington ratio can be  an overestimate of the luminosity caused by gravitational lensing. Allowing for magnification would, at the same time,  justify the apparently small BLR size. 
Moreover, we suggest to take into account  two other concurrent clues. The first concerns the negative correlation between the  $\alpha_{ox}=0.384 \log L_{\nu}(2 keV)/L_{\nu}(2500{\rm\AA}) $   and $L_{\nu}(2500 {\rm\AA})$ found in statistical samples of QSOs/AGNs. With respect to this relation  PG~1247+267, with $\alpha_{ox}=-1.69$  and $L_{\nu}(2500 {\rm\AA})=2.5 \times 10^{47}$ erg s$^{-1}$ \citep{Shem14} deviates from the general trend by an amount which appears significant, despite the relatively large spread around the average relation. Allowing for a gravitational amplification would not change $\alpha_{ox}$ but, changing $L_{\nu}(2500 {\rm\AA})$,  could make this object fully consistent with the general distribution, similarly to the case  of 2XMM J091301.0+525929  which is a confirmed lensed QSO \citep[see][and refs. therein]{Vagn10}. The second clue concerns the \citet{Bald77} effect. With respect to the average negative correlation between the C{\scriptsize IV}  equivalent width (EW) and  \  $\lambda L_{\lambda}(1350 {\rm\AA})$  \citep{Bian12}, again PG~1247+267, with EW=39.5 {\rm\AA} \citep{Shen11}, is deviant and would be brought to full consistency by allowing for  gravitational lensing.  An amplification of about  10 would at the same time: i) account for both these effects; ii) bring  the Eddington ratio to $\sim 1$ and iii) make this object less luminous than S5~0836+071 and consistent with the $\tau_l$ - $L$ relation.

A candidate damped Ly$\alpha$ system (DLA) at $z$=1.223  in the spectrum of PG~1247+267 was analyzed by \citet{Pett99}. 
Though \citet{Turn02} consider the  column density too low to classify this absorption system as DLA, according to the conventional threshold N$_{HI}=2\times 10^{20}$cm$^{-2}$, it could be associated with a foreground lensing galaxy.
 Of course this does not exclude that  anisotropic (close to face-on) emission, possible intrinsic super-Eddington emission, and gravitational lensing occur at the same time. This suggests  to  further observe the spectral variability  to perform velocity resolved RM, and  also to investigate lensing evidences. Finally we remark that PG~1247+267 is the 5-th most luminous of $\sim$ 100,000 QSOs in the \citet{Shen11} catalog. Among the most luminous objects, the fraction of lensed QSOs could be high enough to bias the SE mass estimates and the studies of the  evolution of both the mass function and the Eddington ratio distribution in cosmic time.

\begin{acknowledgements}
We acknowledge funding from PRIN/MIUR-2010 award 2010NHBSBE.
This research is based on observations collected at Copernico  telescope(Asiago, Italy) of the INAF - Osservatorio Astronomico di Padova, and
at Cassini Telescope (Loiano,Italy) of the INAF - Osservatorio Astronomico di Bologna.
\end{acknowledgements}

\bibliography{biblio}{}

\begin{thebibliography}{}
\expandafter\ifx\csname natexlab\endcsname\relax\def\natexlab#1{#1}\fi

\bibitem[{{Alexander}(1997)}]{Alex97}
{Alexander}, T. 1997, in Astrophysics and Space Science Library, Vol. 218,
  Astronomical Time Series, ed. D.~{Maoz}, A.~{Sternberg}, \& E.~M.
  {Leibowitz}, 163

\bibitem[{{Assef} {et~al.}(2011){Assef}, {Denney}, {Kochanek}, {Peterson},
  {Koz{\l}owski}, {Ageorges}, {Barrows}, {Buschkamp}, {Dietrich}, {Falco},
  {Feiz}, {Gemperlein}, {Germeroth}, {Grier}, {Hofmann}, {Juette}, {Khan},
  {Kilic}, {Knierim}, {Laun}, {Lederer}, {Lehmitz}, {Lenzen}, {Mall}, {Madsen},
  {Mandel}, {Martini}, {Mathur}, {Mogren}, {Mueller}, {Naranjo}, {Pasquali},
  {Polsterer}, {Pogge}, {Quirrenbach}, {Seifert}, {Stern}, {Shappee}, {Storz},
  {Van Saders}, {Weiser}, \& {Zhang}}]{Asse11}
{Assef}, R.~J., {Denney}, K.~D., {Kochanek}, C.~S., {et~al.} 2011, \apj, 742,
  93

\bibitem[{{Bachev} {et~al.}(2004){Bachev}, {Marziani}, {Sulentic}, {Zamanov},
  {Calvani}, \& {Dultzin-Hacyan}}]{Bach04}
{Bachev}, R., {Marziani}, P., {Sulentic}, J.~W., {et~al.} 2004, \apj, 617, 171

\bibitem[{Baldwin(1977)}]{Bald77}
Baldwin, J.~A. 1977, ApJ, 214, 679

\bibitem[{{Bentz} {et~al.}(2009){Bentz}, {Peterson}, {Netzer}, {Pogge}, \&
  {Vestergaard}}]{Bent09}
{Bentz}, M.~C., {Peterson}, B.~M., {Netzer}, H., {Pogge}, R.~W., \&
  {Vestergaard}, M. 2009, \apj, 697, 160

\bibitem[{{Bentz} {et~al.}(2006){Bentz}, {Peterson}, {Pogge}, {Vestergaard}, \&
  {Onken}}]{Bent06}
{Bentz}, M.~C., {Peterson}, B.~M., {Pogge}, R.~W., {Vestergaard}, M., \&
  {Onken}, C.~A. 2006, \apj, 644, 133

\bibitem[{{Bentz} {et~al.}(2013){Bentz}, {Denney}, {Grier}, {Barth},
  {Peterson}, {Vestergaard}, {Bennert}, {Canalizo}, {De Rosa}, {Filippenko},
  {Gates}, {Greene}, {Li}, {Malkan}, {Pogge}, {Stern}, {Treu}, \&
  {Woo}}]{Bent13}
{Bentz}, M.~C., {Denney}, K.~D., {Grier}, C.~J., {et~al.} 2013, \apj, 767, 149

\bibitem[{{Bessell}(1990)}]{Bess90}
{Bessell}, M.~S. 1990, \pasp, 102, 1181

\bibitem[{Bian {et~al.}(2012)Bian, Fang, Huang, \& Wang}]{Bian12}
Bian, W.-H., Fang, L.-L., Huang, K.-L., \& Wang, J.-M. 2012, MNRAS, 427, 2881

\bibitem[{{Blandford} \& {McKee}(1982)}]{Blan82}
{Blandford}, R.~D., \& {McKee}, C.~F. 1982, ApJ, 255, 419

\bibitem[{Chelouche {et~al.}(2012)Chelouche, Daniel, \& Kaspi}]{Chel12}
Chelouche, D., Daniel, E., \& Kaspi, S. 2012, ApJ, 750, L43

\bibitem[{{Denney}(2012)}]{Denn12}
{Denney}, K.~D. 2012, ApJ, 759, 44

\bibitem[{Edelson \& Krolik(1988)}]{Edel88}
Edelson, R.~A., \& Krolik, J.~H. 1988, ApJ, 333, 646

\bibitem[{{Ferrarese} \& {Merritt}(2000)}]{Ferr00}
{Ferrarese}, L., \& {Merritt}, D. 2000, \apjl, 539, L9

\bibitem[{Gaskell \& Peterson(1987)}]{Gask87}
Gaskell, C.~M., \& Peterson, B.~M. 1987, ApJS, 65, 1

\bibitem[{{Gaskell C. M.}(2009)}]{Gask09}
{Gaskell C. M.} 2009, New Astronomy Reviews, 53, 140

\bibitem[{{Greene} {et~al.}(2010){Greene}, {Peng}, \& {Ludwig}}]{Gree10}
{Greene}, J.~E., {Peng}, C.~Y., \& {Ludwig}, R.~R. 2010, \apj, 709, 937

\bibitem[{{Kaspi} {et~al.}(2007){Kaspi}, {Brandt}, {Maoz}, {Netzer},
  {Schneider}, \& {Shemmer}}]{Kasp07}
{Kaspi}, S., {Brandt}, W.~N., {Maoz}, D., {et~al.} 2007, \apj, 659, 997

\bibitem[{{Kaspi} {et~al.}(2005){Kaspi}, {Maoz}, {Netzer}, {Peterson},
  {Vestergaard}, \& {Jannuzi}}]{Kasp05}
{Kaspi}, S., {Maoz}, D., {Netzer}, H., {et~al.} 2005, \apj, 629, 61

\bibitem[{{Kaspi} {et~al.}(2000){Kaspi}, {Smith}, {Netzer}, {Maoz}, {Jannuzi},
  \& {Giveon}}]{Kasp00}
{Kaspi}, S., {Smith}, P.~S., {Netzer}, H., {et~al.} 2000, \apj, 533, 631

\bibitem[{{Kelly} {et~al.}(2009){Kelly}, {Bechtold}, \&
  {Siemiginowska}}]{Kell09}
{Kelly}, B.~C., {Bechtold}, J., \& {Siemiginowska}, A. 2009, ApJ, 698, 895

\bibitem[{{Koz{\l}owski} {et~al.}(2010){Koz{\l}owski}, {Kochanek}, {Udalski},
  {Wyrzykowski}, {Soszy{\'n}ski}, {Szyma{\'n}ski}, {Kubiak}, {Pietrzy{\'n}ski},
  {Szewczyk}, {Ulaczyk}, {Poleski}, \& {OGLE Collaboration}}]{Kozo10}
{Koz{\l}owski}, S., {Kochanek}, C.~S., {Udalski}, A., {et~al.} 2010, \apj, 708,
  927

\bibitem[{Krawczyk {et~al.}(2013)Krawczyk, Richards, Mehta, Vogeley, Gallagher,
  Leighly, Ross, \& Schneider}]{Kraw13}
Krawczyk, C.~M., Richards, G.~T., Mehta, S.~S., {et~al.} 2013, ApJS, 206, 4

\bibitem[{{MacLeod} {et~al.}(2010){MacLeod}, {Ivezi{\'c}}, {Kochanek},
  {Koz{\l}owski}, {Kelly}, {Bullock}, {Kimball}, {Sesar}, {Westman}, {Brooks},
  {Gibson}, {Becker}, \& {de Vries}}]{MacL10}
{MacLeod}, C.~L., {Ivezi{\'c}}, {\v Z}., {Kochanek}, C.~S., {et~al.} 2010,
  \apj, 721, 1014

\bibitem[{Marconi {et~al.}(2004)Marconi, Risaliti, Gilli, Hunt, Maiolino, \&
  Salvati}]{Marc04}
Marconi, A., Risaliti, G., Gilli, R., {et~al.} 2004, MNRAS, 351, 169

\bibitem[{Marziani \& Sulentic(2012)}]{Marz12}
Marziani, P., \& Sulentic, J.~W. 2012, New Astronomy Reviews, 56, 49

\bibitem[{{McGill} {et~al.}(2008){McGill}, {Woo}, {Treu}, \& {Malkan}}]{McGi08}
{McGill}, K.~L., {Woo}, J.-H., {Treu}, T., \& {Malkan}, M.~A. 2008, \apj, 673,
  703

\bibitem[{{McIntosh} {et~al.}(1999){McIntosh}, {Rieke}, {Rix}, {Foltz}, \&
  {Weymann}}]{Mcin99}
{McIntosh}, D.~H., {Rieke}, M.~J., {Rix}, H.-W., {Foltz}, C.~B., \& {Weymann},
  R.~J. 1999, \apj, 514, 40

\bibitem[{{McLure} \& {Jarvis}(2002)}]{McLu02}
{McLure}, R.~J., \& {Jarvis}, M.~J. 2002, MNRAS, 337, 109

\bibitem[{{Metzroth} {et~al.}(2006){Metzroth}, {Onken}, \& {Peterson}}]{Metz06}
{Metzroth}, K.~G., {Onken}, C.~A., \& {Peterson}, B.~M. 2006, ApJ, 647, 901

\bibitem[{Netzer(2003)}]{Netz03}
Netzer, H. 2003, ApJ, 583, L5

\bibitem[{{Netzer} {et~al.}(2007){Netzer}, {Lira}, {Trakhtenbrot}, {Shemmer},
  \& {Cury}}]{Netz07}
{Netzer}, H., {Lira}, P., {Trakhtenbrot}, B., {Shemmer}, O., \& {Cury}, I.
  2007, \apj, 671, 1256

\bibitem[{Onken {et~al.}(2004)Onken, Ferrarese, Merritt, Peterson, Pogge,
  Vestergaard, \& Wandel}]{Onke04}
Onken, C.~A., Ferrarese, L., Merritt, D., {et~al.} 2004, ApJ, 615, 645

\bibitem[{{Onken} \& {Peterson}(2002)}]{Onke02}
{Onken}, C.~A., \& {Peterson}, B.~M. 2002, ApJ, 572, 746

\bibitem[{Pancoast {et~al.}(2013)Pancoast, Brewer, Treu, Park, Barth, Bentz, \&
  Woo}]{Panc13}
Pancoast, A., Brewer, B.~J., Treu, T., {et~al.} 2013, arXiv.org, 6475

\bibitem[{{Perna} {et~al.}(2014){Perna}, {Trevese}, {Vagnetti}, \&
  {Saturni}}]{Pern14}
{Perna}, M., {Trevese}, D., {Vagnetti}, F., \& {Saturni}, F.~G. 2014, Advances
  in Space Research, 54, 1429

\bibitem[{{Peterson}(1993)}]{Pete93}
{Peterson}, B.~M. 1993, PASP, 105, 247

\bibitem[{{Peterson} \& {Wandel}(1999)}]{Pete99}
{Peterson}, B.~M., \& {Wandel}, A. 1999, ApJL, 521, L95

\bibitem[{Peterson {et~al.}(1998)Peterson, Wanders, Bertram, Hunley, Pogge, \&
  Wagner}]{Pete98}
Peterson, B.~M., Wanders, I., Bertram, R., {et~al.} 1998, ApJ, 501, 82

\bibitem[{{Peterson} {et~al.}(2004){Peterson}, {Ferrarese}, {Gilbert}, {Kaspi},
  {Malkan}, {Maoz}, {Merritt}, {Netzer}, {Onken}, {Pogge}, {Vestergaard}, \&
  {Wandel}}]{Pete04}
{Peterson}, B.~M., {Ferrarese}, L., {Gilbert}, K.~M., {et~al.} 2004, \apj, 613,
  682

\bibitem[{{Peterson} {et~al.}(2005){Peterson}, {Bentz}, {Desroches},
  {Filippenko}, {Ho}, {Kaspi}, {Laor}, {Maoz}, {Moran}, {Pogge}, \&
  {Quillen}}]{Pete05}
{Peterson}, B.~M., {Bentz}, M.~C., {Desroches}, L.-B., {et~al.} 2005, \apj,
  632, 799

\bibitem[{{Peterson} {et~al.}(2006){Peterson}, {Bentz}, {Desroches},
  {Filippenko}, {Ho}, {Kaspi}, {Laor}, {Maoz}, {Moran}, {Pogge}, \&
  {Quillen}}]{Pete06}
---. 2006, \apj, 641, 638

\bibitem[{Pettini {et~al.}(1999)Pettini, Ellison, Steidel, \& Bowen}]{Pett99}
Pettini, M., Ellison, S.~L., Steidel, C.~C., \& Bowen, D.~V. 1999, ApJ, 510,
  576

\bibitem[{{Pickles} \& {Depagne}(2010)}]{Pick10}
{Pickles}, A., \& {Depagne}, {\'E}. 2010, \pasp, 122, 1437

\bibitem[{{Press} {et~al.}(1992){Press}, {Rybicki}, \& {Hewitt}}]{Pres92}
{Press}, W.~H., {Rybicki}, G.~B., \& {Hewitt}, J.~N. 1992, ApJ, 385, 416

\bibitem[{Press {et~al.}(2007)Press, Teukolsky, Vetterling, \&
  Flannry}]{Pres07}
Press, W.~H., Teukolsky, S.~A., Vetterling, W.~T., \& Flannry, B.~P. 2007,
  Numerical Recipes, The Art of Scientific Computing, Third Edition (Cambridge
  University Press), 824

\bibitem[{{Ramsey} {et~al.}(1998){Ramsey}, {Adams}, {Barnes}, {Booth},
  {Cornell}, {Fowler}, {Gaffney}, {Glaspey}, {Good}, {Hill}, {Kelton},
  {Krabbendam}, {Long}, {MacQueen}, {Ray}, {Ricklefs}, {Sage}, {Sebring},
  {Spiesman}, \& {Steiner}}]{Rams98}
{Ramsey}, L.~W., {Adams}, M.~T., {Barnes}, T.~G., {et~al.} 1998, in Society of
  Photo-Optical Instrumentation Engineers (SPIE) Conference Series, Vol. 3352,
  Advanced Technology Optical/IR Telescopes VI, ed. L.~M. {Stepp}, 34--42

\bibitem[{Runnoe {et~al.}(2014)Runnoe, Brotherton, DiPompeo, \& Shang}]{Runn14}
Runnoe, J.~C., Brotherton, M.~S., DiPompeo, M.~A., \& Shang, Z. 2014, MNRAS,
  438, 3263

\bibitem[{{Runnoe} {et~al.}(2013){Runnoe}, {Brotherton}, {Shang}, \&
  {DiPompeo}}]{Runn13}
{Runnoe}, J.~C., {Brotherton}, M.~S., {Shang}, Z., \& {DiPompeo}, M.~A. 2013,
  \mnras, 434, 848

\bibitem[{{Rybicki} \& {Kleyna}(1994)}]{Rybi94}
{Rybicki}, G.~B., \& {Kleyna}, J.~T. 1994, in Astronomical Society of the
  Pacific Conference Series, Vol.~69, Reverberation Mapping of the Broad-Line
  Region in Active Galactic Nuclei, ed. P.~M. {Gondhalekar}, K.~{Horne}, \&
  B.~M. {Peterson}, 85

\bibitem[{{Rybicki} \& {Press}(1992)}]{Rybi92}
{Rybicki}, G.~B., \& {Press}, W.~H. 1992, ApJ, 398, 169

\bibitem[{{Saturni} {et~al.}(2014){Saturni}, {Trevese}, {Vagnetti}, \&
  {Perna}}]{Satu14}
{Saturni}, F.~G., {Trevese}, D., {Vagnetti}, F., \& {Perna}, M. 2014, Advances
  in Space Research, 54, 1434

\bibitem[{Shemmer {et~al.}(2014)Shemmer, Brandt, Paolillo, Kaspi, Vignali,
  Stein, Lira, Schneider, \& Gibson}]{Shem14}
Shemmer, O., Brandt, W.~N., Paolillo, M., {et~al.} 2014, ApJ, 783, 116

\bibitem[{Shen \& Kelly(2012)}]{Shen12}
Shen, Y., \& Kelly, B.~C. 2012, ApJ, 746, 169

\bibitem[{{Shen} {et~al.}(2011){Shen}, {Richards}, {Strauss}, {Hall},
  {Schneider}, {Snedden}, {Bizyaev}, {Brewington}, {Malanushenko},
  {Malanushenko}, {Oravetz}, {Pan}, \& {Simmons}}]{Shen11}
{Shen}, Y., {Richards}, G.~T., {Strauss}, M.~A., {et~al.} 2011, \apjs, 194, 45

\bibitem[{Sluse {et~al.}(2011)Sluse, Schmidt, Courbin, Hutsem{\'e}kers, Meylan,
  Eigenbrod, Anguita, Agol, \& Wambsganss}]{Slus11}
Sluse, D., Schmidt, R., Courbin, F., {et~al.} 2011, Astronomy and Astrophysics,
  528, 100

\bibitem[{{Sulentic} {et~al.}(2007){Sulentic}, {Bachev}, {Marziani}, {Negrete},
  \& {Dultzin}}]{Sule07}
{Sulentic}, J.~W., {Bachev}, R., {Marziani}, P., {Negrete}, C.~A., \&
  {Dultzin}, D. 2007, \apj, 666, 757

\bibitem[{{Sulentic} {et~al.}(2002){Sulentic}, {Marziani}, {Zamanov}, {Bachev},
  {Calvani}, \& {Dultzin-Hacyan}}]{Sule02}
{Sulentic}, J.~W., {Marziani}, P., {Zamanov}, R., {et~al.} 2002, \apjl, 566,
  L71

\bibitem[{{Trevese} {et~al.}(2007){Trevese}, {Paris}, {Stirpe}, {Vagnetti}, \&
  {Zitelli}}]{Trev07}
{Trevese}, D., {Paris}, D., {Stirpe}, G.~M., {Vagnetti}, F., \& {Zitelli}, V.
  2007, \aap, 470, 491

\bibitem[{{Trevese} {et~al.}(2013){Trevese}, {Saturni}, {Vagnetti}, {Perna},
  {Paris}, \& {Turriziani}}]{Trev13}
{Trevese}, D., {Saturni}, F.~G., {Vagnetti}, F., {et~al.} 2013, \aap, 557, A91

\bibitem[{Turnshek \& Rao(2002)}]{Turn02}
Turnshek, D.~A., \& Rao, S.~M. 2002, ApJ, 572, L7

\bibitem[{{Vagnetti} {et~al.}(2010){Vagnetti}, {Turriziani}, {Trevese}, \&
  {Antonucci}}]{Vagn10}
{Vagnetti}, F., {Turriziani}, S., {Trevese}, D., \& {Antonucci}, M. 2010, \aap,
  519, A17

\bibitem[{Vestergaard(2002)}]{Vest02}
Vestergaard, M. 2002, ApJ, 571, 733

\bibitem[{{Vestergaard}(2004)}]{Vest04}
{Vestergaard}, M. 2004, \apj, 601, 676

\bibitem[{{Vestergaard} \& {Peterson}(2006)}]{Vest06}
{Vestergaard}, M., \& {Peterson}, B.~M. 2006, ApJ, 641, 689

\bibitem[{Wandel {et~al.}(1999)Wandel, Peterson, \& Malkan}]{Wand99}
Wandel, A., Peterson, B.~M., \& Malkan, M.~A. 1999, ApJ, 526, 579

\bibitem[{{White} \& {Peterson}(1994)}]{Whit94}
{White}, R.~J., \& {Peterson}, B.~M. 1994, PASP, 106, 879

\bibitem[{Zu {et~al.}(2013)Zu, Kochanek, Koz{\l}owski, \& Peterson}]{Zu13b}
Zu, Y., Kochanek, C.~S., Koz{\l}owski, S., \& Peterson, B.~M. 2013, arXiv.org,
  6774

\bibitem[{{Zu} {et~al.}(2013){Zu}, {Kochanek}, {Koz{\l}owski}, \&
  {Udalski}}]{Zu13a}
{Zu}, Y., {Kochanek}, C.~S., {Koz{\l}owski}, S., \& {Udalski}, A. 2013, \apj,
  765, 106

\bibitem[{{Zu} {et~al.}(2011){Zu}, {Kochanek}, \& {Peterson}}]{Zu11}
{Zu}, Y., {Kochanek}, C.~S., \& {Peterson}, B.~M. 2011, ApJ, 735, 80

\end{thebibliography}
\end{document}